\documentclass[final,12pt]{elsarticle}

\usepackage{amssymb}

\journal{Astroparticle Physics}

\begin{document}

\begin{frontmatter}



\title{MACHETE: A transit Imaging Atmospheric Cherenkov Telescope to 
survey half of the Very High Energy $\gamma$-ray sky}


\author[label1]{J. Cortina}
\ead{Corresponding author: cortina@ifae.es}
\author[label1]{R. L\'opez-Coto}
\author[label1]{A. Moralejo}

\address[label1]{Institut de Fisica d'Altes Energies\\
Edifici CN, Campus UAB\\
E-08193, Cerdanyola del Valles, Spain}

\begin{abstract}
Current Imaging Atmospheric Cherenkov Telescopes for Very High Energy $\gamma$-ray astrophysics 
are pointing instruments with a Field of View up to a few tens of sq~deg. We propose to build an 
array of two non-steerable (drift) telescopes. Each of the telescopes would have a 
camera with a FOV of 5$\times$60~sq~deg oriented along the meridian. 
About half of the sky drifts through this FOV in a year. 
We have performed a Montecarlo simulation to estimate the performance of this instrument.
We expect it to survey this half of the sky with an integral flux sensitivity of 
$\sim$0.77\% of the steady flux of the Crab Nebula
in 5 years, an analysis energy threshold of $\sim$150 GeV and an 
angular resolution of $\sim$0.1$^{\circ}$. For astronomical objects that transit over the 
telescope for a specific night, we can achieve an integral sensitivity of 12\% of the Crab Nebula 
flux in a night, making it a very powerful tool to trigger further observations of variable sources 
using steerable IACTs or instruments at other wavelengths.

\end{abstract}


\end{frontmatter}




\section{Introduction}
\label{Introduction}

Very High Energy (VHE; $>$100~GeV) $\gamma$-rays are detected using space-based or ground-based 
detectors. From space {\it Fermi}-LAT\cite{fermi} is performing the deepest survey to date 
of the $\gamma$-ray sky from 20~MeV up to energies in excess of 100~GeV, although with limited 
sensitivity above 10~GeV due to its relatively small collection area (0.8~m$^2$). 

From the ground
Imaging Atmospheric Cherenkov Telescopes (IACTs), such as the MAGIC\cite{magic1,magic2}, 
H.E.S.S.\cite{hess} or VERITAS\cite{veritas} arrays, detect
$\gamma$-rays with energies above 50~GeV and have collection areas of more than 10$^5$~m$^2$. They 
are pointing instruments with a Field of View (FOV) on the order of tens of sq~deg. 
The 12~m diameter Medium-Sized Telescopes in the Cherenkov Telescope Array (CTA), currently 
under design, have a FOV of around 60~sq~deg. 
The telescope mirrors have an area on the order of hundreds of square meters. 

On the other hand air-shower instruments such as Milagro\cite{milagro}, Tibet\cite{tibet} and 
HAWC\cite{hawc} detect $\gamma$-rays at higher energies 
(above some hundreds of GeV to typically TeV), have a comparable collection area of 
$\sim$80000~m$^2$, but 
with a much larger FOV of $\sim$5000 sq~deg and high duty cycle. They are non-tracking instruments. 
Unfortunately they are not as efficient as IACTs in eliminating the cosmic ray background, 
so they suffer from a lower sensitivity and they have poorer angular or spectral resolutions.

We propose to build an array of two non-steerable IACTs with a wide FOV of 300 sq~deg. 
We call this array Meridian Atmospheric CHErenkov TElescope (MACHETE). MACHETE's 
FOV is significantly larger than
that of all existing IACTs, making it into an instrument suitable for surveys. On the other hand 
its sensitivity is as good as that of current IACT arrays.
In the next sections we describe the instrument,
we evaluate its expected performance and we discuss its main physics goals.

\section{Telescope design}

\subsection{Basics of the optical design}
\label{Optics}

Steerable IACTs with wide FOV of tens of sq~deg have been proposed 
\cite{schliesser,vassiliev,maccarone,mirzoyan}. 
Ultra High Energy cosmic ray detectors using the fluorescence technique with a very 
large FOV on the order of steradians are already operational\cite{auger} or under design\cite{euso}, 
but their light collection area does not exceed a few square meters. 

A Schmidt telescope is a well-known solution to achieve a large FOV with a small focal ratio. 
The optical components are an easy-to-manufacture spherical primary mirror, and an 
aspherical correcting lens, 
known as a Schmidt corrector plate, located at the center of curvature of the primary mirror.
The corrector plate reduces optical aberrations and at the same time acts as a ``stop'' which 
defines the aperture of the telescope. The authors of \cite{mirzoyan} have in fact proposed a 
Schmidt IACT with a 7~m diameter primary mirror. 
Their design is technically challenging though, because the corrector plate is as large as 7~m and 
it is implemented as a thin tesselated Fresnel lens.

Our concept is inspired by a Schmidt telescope, but we have aimed at simplifying it so that it is
easy and cheap to implement:
\begin{itemize}
\item Like in the original Schmidt telescope, the shape of the primary mirror is spherical. 
The nominal focal length is half of the radius of curvature.
\item Also like a Schmidt telescope, the shape of the focal plane is spherical
and concentric with the mirror.
\item An IACT is not as stringent as an optical telescope in terms of mirror 
Point Spread Function (PSF). A PSF on the order of 0.05$^{\circ}$ is good enough.
We shall remove the corrector plate and achieve an acceptable PSF by increasing 
the focal ratio. 
\item However if we eliminate the corrector plate we are not only worsening the
optical performance of the instrument, but also eliminating the stop. We must find an
alternative way to limit the aperture.
Compared to optical telescopes IACTs are in fact peculiar because each pixel is typically
implemented as a light concentrator, usually a Winston Cone (WC), followed by the 
actual photodetector. The light concentrator serves three purposes: it limits stray 
light beyond a certain acceptance angle and it allows to reduce the dead space and the 
size of the photodetector (and hence its cost). But in a natural way 
light concentrators can also be used to define the section of the mirror which is viewed 
by each pixel and effectively the aperture. 
\end{itemize}

For MACHETE we adopt the following optical parameters. The radius of curvature of the mirror
is 34~m. We choose an acceptance angle of 20$^{\circ}$ in the light concentrators. 
The light concentrators at the camera front follow the curvature of the focal plane.
The focal plane is at roughly half of the radius of curvature (17~m), so each point of the camera 
views a section of the mirror that is circular and has a diameter D=12~m. This means that 
the aperture becomes effectively 12~m and the focal ratio is f/D=1.42.

\begin{figure}[!t]
  \vspace{5mm} \centering
  \includegraphics[height=13cm]{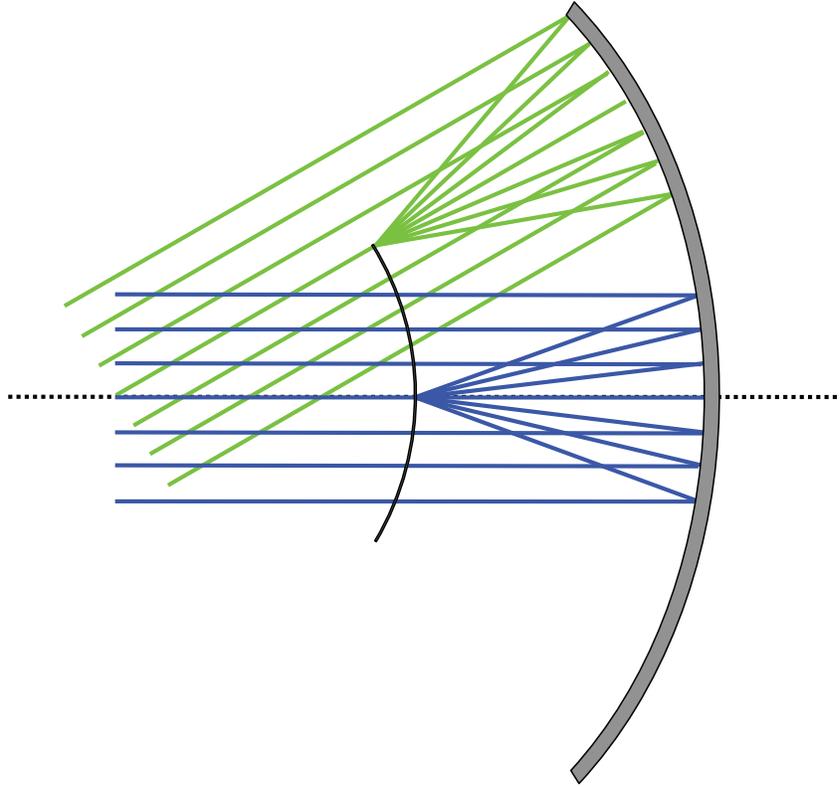}
  \caption{\small{Layout of the optical elements of MACHETE. The dotted line represents the 
      optical axis. Both mirror (outer grey arc) and focal plane (inner black arc) are
      concentric. Rays coming parallel to the optical axis and 30$^{\circ}$ off-axis are
      represented with correspondingly blue and green lines.}}
  \label{fig:layout}
\end{figure}
Figure \ref{fig:layout} illustrates the concept. We have drawn the optical path of a fan of 
rays coming parallel to the optical axis and of a fan of rays with a large 30$^{\circ}$ off-axis angle. 
The central ray of each of the fans goes through the center of curvature of the mirror 
and focal plane. As such it arrives
perpendicular to both surfaces. The light concentrators limit the
extension of the ray fans in figure \ref{fig:layout} and the effective diameter of the mirror that
collects light for every point in the camera.

We have used Zemax (OpticStudio 14.2) to optimize the optical layout of the design. We set 
the radius of curvature of the mirror to 34~m. We have not simulated the light concentrators. 
Instead we have defined a 12~m diameter circular stop perpendicular to the incident ray fan 
and centered at the center of the curvature of both mirror and focal plane. Both scenarios
are optically interchangeable if the light concentrators have a ideally sharp cutoff.

The distance from the mirror to the camera front
has been optimized to obtain the smallest possible spot size. More specifically
we have defined r$_{80}$ as the radius around the centroid of the light spot 
that encloses 80\% of the light and we have minimized r$_{80}$ on-axis. The resulting distance from mirror to
camera front is 16.84~m and the radius of curvature of the camera is correspondingly 17.16~m. 
The plate scale is 300~mm/deg.

\begin{figure}[!t]
  \vspace{5mm} \centering
  \includegraphics[height=8cm]{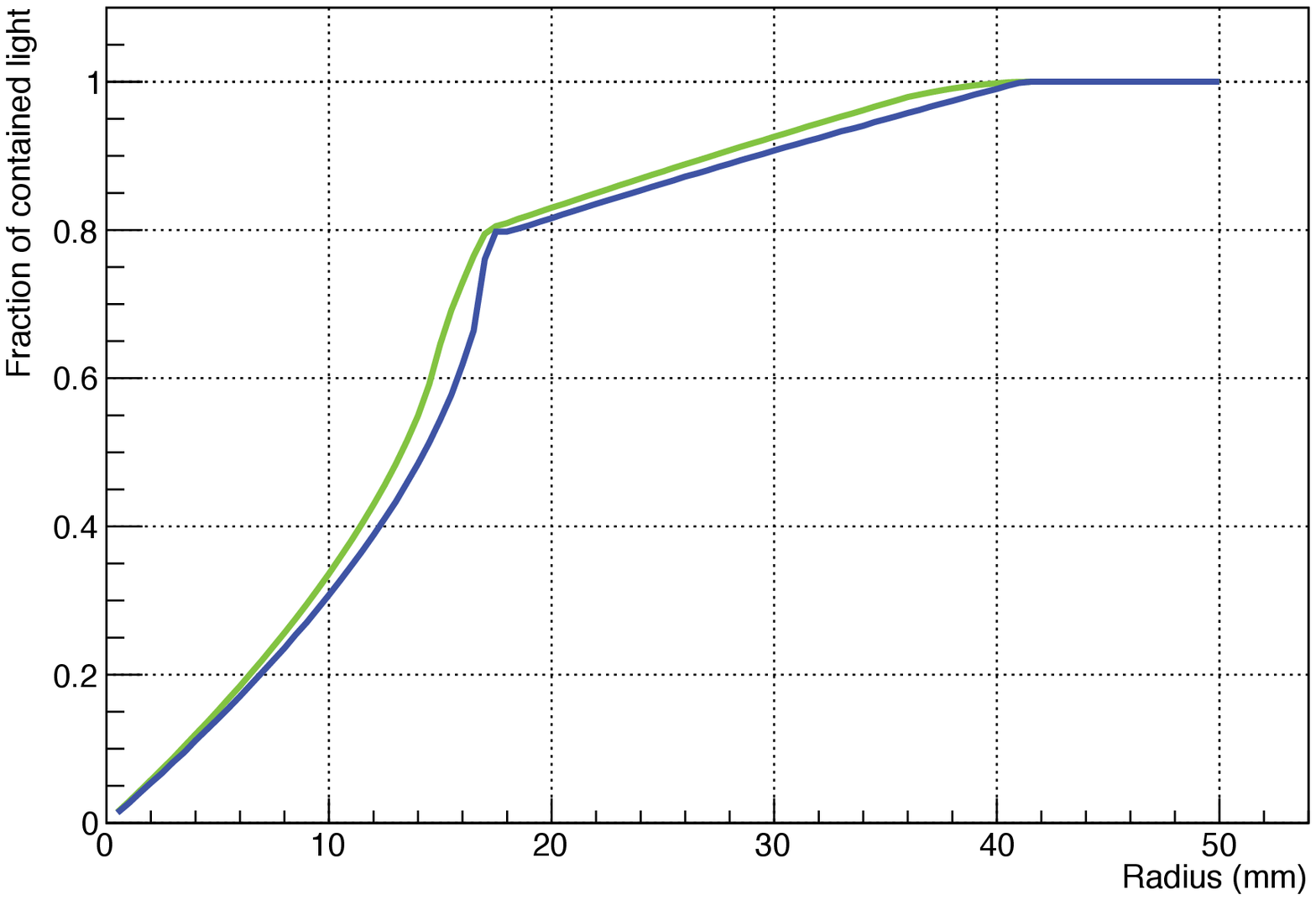}
  \caption{\small{Fraction of the incident light that is focused inside a given radius around 
      the spot centroid as a function of that radius. The curves correspond to light incident 
      along the optical axis (blue) and light incident 30$^{\circ}$ off-axis (green). There is 
      hardly any difference between the two curves: r$_{80}$ is 17.5~mm for both of them. The 
      change of slope around 17.5~mm is related to the fact that the camera is not located 
      at the nominal focal distance of the spherical mirror.}}
  \label{fig:encircled_energy}
\end{figure}
Figure \ref{fig:encircled_energy} shows the fraction of the incident light that is enclosed within a 
certain radius around the spot centroid for two angles respect to the optical axis. This 
fraction has been calculated using Zemax. r$_{80}$ is 17.5~mm for both incident angles, which 
corresponds to 0.06$^{\circ}$. As may be expected there is hardly any difference in r$_{80}$ 
for any position of the FOV.

This Zemax simulation does not take into account the effect of the spread in the distribution of
actual radii of curvature of the facets during the fabrication process or the alignment of the 
facets during the installation process. We will come back to the error introduced by these two 
effects in section \ref{Focal}.

With a camera as large as in figure \ref{fig:layout} all rays coming parallel to the 
optical axis are blocked if we assume that the system is symmetric. The rest of the FOV would 
also suffer from significant shadowing, i.e. the instrument
will suffer from strong vignetting. We can however restrict our FOV to a strip of 
5$^{\circ}$$\times$60$^{\circ}$.
Assuming that the focal plane instruments are flat enough and ignoring the focal plane support 
structure, the shadowing is 16\% for most of the FOV 
and only significantly smaller at the edges of the long arc (at the very edge it is about 8\%).
In this way we can still achieve a large FOV of 300~sq~deg with a very simple optical design.

In a spherical reflector photons from the same direction but hitting the reflector at different 
distances to the optical axis arrive to the focus at different times. For this specific optical 
setup, the largest time difference is 3.7~ns and the RMS of the distribution is about 1.5~ns. 
This time spread hardly degrades the time resolution of the pixel signal reconstruction 
although it may worsen the trigger threshold of the
telescope. We will evaluate this effect more carefully in section \ref{Performance}.

\subsection{Orientation of the telescope}
\label{Orientation}

IACTs are usually steerable. This allows to observe a source for many hours in a night
and up to hundreds of hours in a year. However the fact that the reflector must be steerable 
limits the weight of camera and reflector, makes motors, drives and other steering and monitoring 
elements necessary, and complicates the alignment of the reflector mirrors.
Since our aim is to survey a significant fraction of the sky, there is no need to track sources.
We can point the telescope to an arbitrary direction and wait for sources to drift through the FOV.

IACTs achieve their best sensitivity and lowest energy threshold at small zenith angles 
($<$30$^{\circ}$) so it is optimal to observe sources during culmination. 
We propose to place our optical axis vertical and align the FOV with the meridian. 
In this way the telescope will only have access to astronomical objects in the declination 
range of $\pm$30$^{\circ}$ 
around the geographical latitude where it is located and only for about 20~minutes every night 
as they culminate for that specific geographical longitude.  

\begin{figure}[!t]
  \vspace{5mm} \centering
  \includegraphics[height=8cm]{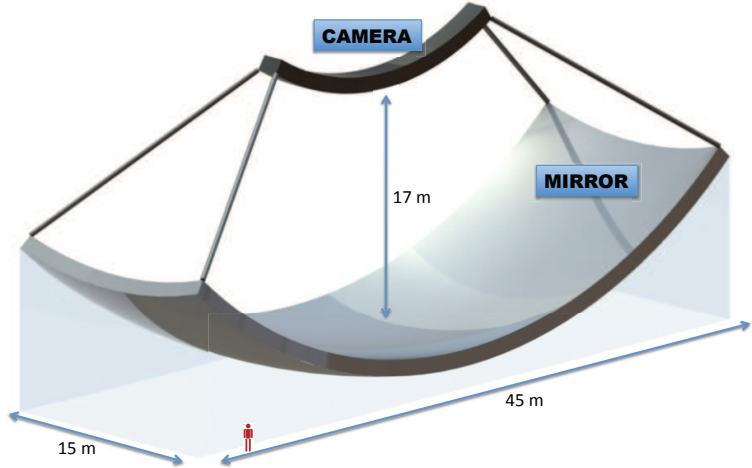}
  \caption{\small{Conceptual design of one of the two telescopes. 
      The mirror is a rectangular section of a spherical mirror, is fixed to the ground 
      and its long axis is oriented north to south. The camera has a spherical curvature, 
      is concentric with the mirror and has about half of its radius of curvature. Its FOV 
      is a rectangle of 5$^{\circ}$$\times$60$^{\circ}$. A human figure has been
      added as a size reference.}}
  \label{fig:mechanical}
\end{figure}
We will extend the mirror so that all points in the FOV view a circular fraction of the total 
reflector with a diameter of 12~m. For the above-mentioned focal length, this corresponds to a total 
reflective surface of 620~m$^2$. From north to south the reflector has a length of about 45~m, 
from east to west it has a width of about 14~m.
The physical size of the focal plane is about 27~m$^2$. It has a width of 1.5~m and a length 
(along the arc) of 18~m. Figure \ref{fig:mechanical} is a conceptual view of such an IACT. 

The reflector shall be tesselated. Each of the facets would have a spherical shape with
the same radius of curvature as the global shape of the reflector and would be arranged 
following the shape of the global sphere. 

IACTs are actually not focused to infinity but to the average position of the atmospheric
shower which is generated by the $\gamma$-ray for the energy of interest. Typically this position 
is about 10 km away from the telescopes. This makes an additional shift of 3 cm of the focal plane 
in the direction of the mirror necessary (along with some adjustment of the mirror facets if the
reflector is not spherical). This distance depends on zenith angle. 
For a steerable IACT one needs to shift the whole camera for a distance which depends on the zenith 
angle of the observation. In our optical setup however each point of the FOV (each pixel) 
corresponds to only one zenith angle, so one may 
shift each individual pixel to a fixed optimal position with no need for dynamic adjustment.

In order to benefit from stereoscopic reconstruction of atmospheric showers we need to work 
with at least two telescopes. We propose to build two identical telescopes with the 
above-described optical design and orientation at a distance of $\sim$100~m in the east-west 
direction. This is a compromise between optimal stereo reconstruction and rate of events, but it 
has not been optimized for our setup. We have selected a value comparable to 
telescopes of similar threshold like VERITAS or H.E.S.S.

\section{Focal plane instrumentation and readout}
\label{Focal}

Designing the focal plane instrumentation and readout system of MACHETE goes beyond the scope of
this paper. We will only define a few parameters that are essential to determine the 
performance of the instrument. In general we assume that the performance of these hardware components is 
similar to those in the MAGIC array. The final design would be cost-optimized.

In IACTs it is usual to define the radius of the pixels so that they enclose about 80\% of the light.
For the optics described in the previous section 80\% of the light is contained in a circle of radius
r$_{80}$= 0.06$^{\circ}$. We will assume that the reflector is tesselated with mirror facets of 
around 1~m$^2$. The precision in producing and aligning the facets will, conservatively, 
increase r$_{80}$ to 0.08$^{\circ}$. In order to cover a FOV of 5$^{\circ}$$\times$60$^{\circ}$ 
with pixels of a diameter 0.16$^{\circ}$ we need about 15000 pixels in each telescope. 

Progress in photodetectors has been significant during the last years. For instance the 
photomultipliers (PMTs) that will be installed in the prototype Large-Sized Telescope (LST) of CTA 
have a peak Quantum Efficiency (QE) of $\sim$42\%\cite{cta_pmts} to be compared with a peak 
QE of the MAGIC PMTs of $\sim$32\%\cite{magic1}. On the other hand Silicon PMs (SiPMs) 
reach Photodetection Efficiencies (PDE) in excess of 45\%\cite{icrc_sipms}.

SiPMs have drawbacks when operating in IACTs: they have low detection efficiency at near UV, 
high dark count rate and optical crosstalk, too high sensitivity at long wavelengths where 
Night Sky Background 
(NSB) dominates, and are not currently available for a pixel size in excess of $\sim$20 mm. 
However coupling several small SiPMs into a large detector is possible and the detector
noise hardly ever dominates over NSB. Furthermore progress in the field of SiPM is quick and
FACT\cite{Fact} has already proved that SiPM can be used in IACTs. Advantages of SiPM are 
their high PDE, no ageing effects (after 3 years of operation none of the SiPMs in FACT has 
failed or shown any indication of ageing), excellent uniformity and, relevant for MACHETE, 
the fact that they can be used under the brightest Moonlight.

We will not select a specific type of photosensor though. We will just assume that the 
photodetectors of MACHETE have a PDE 1.5 times larger than the PMTs 
in operation in MAGIC but the same performance in terms of noise, spectral dependence of QE and 
time response. 

In what follows we will assume that the rest of the electronic readout chain has the same features and 
performance as MAGIC. Let us anyway discuss some possible technical solutions. 

The angular extension of a typical 
$\gamma$-ray image is $<$2$^{\circ}$. This is much smaller
than the total FOV, so it makes sense to trigger and read out only a limited part of the camera.
The readout's Region of Interest (RoI) may be as small as 500~pixels around the shower image. 
The readout should be designed to exploit this fact, hence to consider only channels in the RoI beyond a 
certain processing stage.

A local trigger will be based on a time coincidence of a few pixels in a compact cluster
in one of the two telescopes. A stereo trigger will require a time coincidence of a local
trigger in the two telescopes. A possible solution\cite{Paoletti} may be to split the camera 
in 10-20 ``sub-cameras'', to digitize only channels in sub-cameras where a local trigger has been
generated and to store the digital data only if a stereo trigger arrives within a certain time interval. 

However this requires to equip every pixel with a digitizer.
A more speculative solution could be to use the local trigger to activate a fast ``electronic switch'' 
which may forward only signals in the RoI to 500~digitizers (see \cite{mux} for a similar solution
using an optical switch). This would represent a significant cost 
reduction for readout and data storage, but may be technically challenging, for instance due to 
an increase of deadtime.

\section{Some words on the cost of MACHETE}
\label{Cost}

MACHETE has features that make the operation and construction easier and less expensive 
than those of a steerable IACT. Let us go through some of them.

To begin with, camera and reflector do not move. This has numerous advantages:

\begin{itemize}
\item The alignment of the mirror facets is simplified. Considering that the reflector 
does not move and that it 
can be shielded from the wind, once the facets are aligned they are probably stable for 
years. 
There is no need for an active mirror adjustment as implemented in MAGIC.
We may also consider aligning the facets using screws (as it is already done in VERITAS) 
and not mechanical actuators as it is done in many of the IACTs. We also require a much less 
demanding system to monitor the optical PSF and telescope pointing. 
\item In steerable IACTs the weight of mirrors and focal plane instrumentation must be strongly limited to 
reduce the demands on the steering system and to reduce deformations of the mechanical structure.
In our design these limits can be relaxed. The cost of the mirrors can be correspondingly reduced, maybe
to as much as 1000~EUR/m$^2$\cite{Doro}. The camera may get heavier as well and may be afford more efficient
cooling systems, which maybe necessary in case we used SiPMs.
\item Mechanics are more simple. There is no need for azimuth or zenith 
steering: no motors, gears etc. In turn the peak power consumption demands of the telescope, which 
are typically dominated by fast repositioning, are much lower. 
\item Steerable IACTs are parked looking away from the Sun during 
the day or whenever they are not operating, so that the plane of the reflector is essentially vertical. This
makes them especially sensitive to the wind and imposes extra requirements to mechanical supports and 
foundations.
In our instrument the reflector points vertical and it can be shielded from the wind by walls.
\end{itemize}

Secondly, even if a camera is equipped with 15000 pixels, we only need to store data for a RoI of about 
500~pixels around the shower image (see section \ref{Focal}). The cost reduction factor depends
on the actual technical solution implemented. It must be noted that the cost of the readout
may be much lower than the cost of the photodetector with a digitizing chip such as TARGET5\cite{target}.

Finally it is worth to mention that all mirror facets have exactly the same curvature in this optical design. 
This facilitates production, installation and alignment.

On the negative side IACTs are not protected with domes, because the size of the reflectors makes
them prohibitely expensive, but conventional IACTs can be moved away from the Sun during the day.
This is not the case for MACHETE. The reflector (or a large fraction of it) must be covered during the day.
A simple low-weight roof following the shape of the mirror should however suffice. 

More relevant is the cost of the telescope cameras. Each of them needs to be equipped with 15000 
photodetectors. As of today IACT cameras are equipped with photomultipliers and the resulting cost 
would by far dominate over the cost of the rest of the instrument. 

Using SiPMs as photodetectors will foreseeably bring the cost down in the near 
future, maybe to as low as 1~US\$/mm$^2$. 
As mentioned above each pixel must be equipped with a WC to effectively limit the mirror diameter.
Each WC has a hexagonal entrance window of 45~mm side to side and a circular exit window of 
around 15~mm diameter. This means that only one ninth of the focal plane total area must 
be instrumented with photodetectors.
Even so the total cost of the photodetectors for one telescope would be on the order of 2.5 million US\$.

\section{Expected $\gamma$-ray detection performance}
\label{Performance}

\subsection{Setup of the Montecarlo simulation}

We have performed a Monte Carlo simulation of the MACHETE system to
evaluate its performance. The performance of MACHETE anywhere within
its FOV should be the same except for camera edge effects,
therefore a simulation of a square section of the MACHETE cameras,
with the corresponding part of the mirror dish, is enough to
assess the performance of the system as a whole. We have simulated a
system of two such telescopes, at a distance of 85 m\footnote{Actually
we have used as relative telescope positions those of the MAGIC
telescopes, i.e. along a line roughly in the NW - SE direction. A more
realistic configuration would have the two telescopes placed on a line
along the E-W direction, but given that the zenith angle range covered
by MACHETE is 0-30$^\circ$, no significant effect on performance is expected.}. 
We have opted for such a simplified simulation, which can be carried
out with the same tools employed in the simulation of conventional IACTs. 

For the simulation of atmospheric showers we use the CORSIKA (COsmic
Ray SImulations for KAscade) 6.019 program 
\cite{CORSIKA}. The model to describe hadronic interactions is
FLUKA, together with the  QGSJet-II model for high-energy
interactions. Electromagnetic interactions are simulated using the
EGS4 model.  The detector simulation (optical system and electronics)
and the analysis of the simulated events have been carried out using
the official software of the MAGIC collaboration\cite{magic_sw0,magic_sw1,magic_sw2}. 
The simulated telescopes were placed at the MAGIC site, i.e. at an elevation of
2200 m above sea level.

The simulated primaries are:
\begin{itemize}
\item {\bf Gamma rays: } They are simulated with a power-law
  differential energy spectrum with index -2.0 in the range from 10
  GeV to 30 TeV. Gamma rays are simulated as coming from a
  point-like  source, and with impact parameters up to 300 m from the
  center of the array.  The incident directions span the range up to
  35$^\circ$ in zenith distance. A total of $5 \times 10^6$  gamma
  rays were simulated.  
\item {\bf Protons: } They are simulated with a power-law energy
  spectrum  with index -2.0, between 70 GeV and 30 TeV. The 
  maximum impact parameter with respect to the center of the array is
  500 m. The pointing directions also span the range up to 35$^\circ$
  in zenith distance. We simulated $1.3 \times 10^6$ proton showers, each of
  which was processed 20 times through the detector simulation using
  different, random telescope pointing directions within a 5-degree
  semi-aperture cone around the proton direction, to simulate an
  isotropic flux. Since the probability of any given proton to survive
  the trigger and the whole of the analysis chain is very small, this
  recycling of events does not result in strong correlations among the
  events in the final sample. About $8.7 \times 10^6$ proton events passed 
  the trigger.

\end{itemize}
No electron primaries were simulated: in order to calculate their
contribution to the isotropic background, the response of MACHETE to
cosmic ray electrons is assumed to be roughly the same as for gamma
rays. The contribution from showers initiated by  nuclei heavier than
protons, after all cuts, is estimated roughly as a 20\%  of that from
protons, which is added to the estimated proton rates in the
sensitivity calculations below.

The background light of the night sky is simulated at the level found
in typical extragalactic fields observed from the MAGIC site. 

Once atmospheric showers have been simulated and the information of
the Cherenkov photons reaching the observation level is recorded, 
we simulate the response of the reflector dishes. This part of the
simulation and the one described in next section use the software
developed by the MAGIC collaboration to simulate the telescope
response. 
The simulated 12-m diameter mirror dishes are tessellated in squared
panels of  1 m$\times$1 m. Each panel has a spherical surface with
focal distance 1700 cm. Mirror imperfections and tolerances in mirror
alignment similar to those of MAGIC are simulated, resulting in a
worsening of the ideal PSF of the whole dish from 0.06$^\circ$ to
0.08$^\circ (r_{80})$. The camera is located at the distance providing
the optimal PSF for focusing light emitted 10 km above the telescopes.

The \texttt{reflector} program simulates the attenuation of Cherenkov
light in the atmosphere and calculates the position and arrival time
of the photons in the camera plane. The program allows to point the
telescopes in slightly different directions with respect to the shower
axes, so that we can evaluate the off-axis performance. For gamma
rays, discrete off-axis angles of 0.4$^\circ$, 1.0$^\circ$, 1.5$^\circ$
and 2.0$^\circ$ have been simulated . 

The simulated camera is a square with a size of
4$^\circ\times4^\circ$, hence narrower than the smaller of the MACHETE
camera sides ($5^\circ$), but sufficient to evaluate the response of
the full MACHETE, both for sources well contained in the FOV (we
expect a flat response near the center) and for sources close to or
beyond the camera edges. Note that the MACHETE optical
performance is similar over the whole FOV. Additionally, on a
small section of the camera like the simulated one, the curvature of
the focal plane can be (and has been) neglected.

For the full $5^\circ \times 60^\circ$ MACHETE cameras, the performance is 
extrapolated from that of the simplified simulation, taking care to
account for the expected differences in edge effects (which would
affect a significantly smaller fraction of events in the full MACHETE
cameras). The simulated camera is composed of 739 circular
photodetectors attached to WCs, on a hexagonal pattern of
pixels of 0.16$^\circ$ size (flat-to-flat). WCs with a hexagonal entrance 
window do not have a well-defined cutoff angle, so the light concentrator will
need to be further optimized in the real detector. In the simulation 
we have assumed an average light transmission efficiency of 90\%, with no 
angular cutoff.

The \texttt{camera} program simulates the response of the
photodetectors, the trigger system and the readout electronics. For the
photodetectors we have scaled up their QE [MACHETE photodetectors]=
1.5$\cdot$QE [MAGIC PMTs], but we have maintained the same wavelength dependence. 

The system is triggered if there are 3 pixels in close-compact
configuration (each of them next to the other two) above a
discriminator threshold of 5.7 phe in a time shorter than 5.5 ns.
Since we are using a higher QE and larger pixel sizes than in MAGIC,
we had to increase the discriminator threshold with respect to that
used by MAGIC to avoid too many triggers due to the NSB level. 
The single-telescope trigger rate was set to 16 kHz (dominated by NSB
triggers) for the simulated camera of 4$^\circ\times4^\circ$, which is
similar to the current L1 rate of MAGIC. This would result in a much
larger single-telescope accidental rate (by a factor $\sim$20) for the whole FOV of
MACHETE. However the data acquisition would be triggered only by events 
firing both MACHETE telescopes (``stereo trigger''). Assuming a simple time 
coincidence stereo trigger with a coincidence window of 200 ns, the expected
accidental rate for the whole FOV of MACHETE is 900 Hz. In fact the 
stereo rate is dominated by cosmic ray triggers, which produce a rate
as high as 9 kHz according to the simulation.
Whether this may result in a significant dead time fraction,
depends on the details of the trigger and readout implementation,
and is beyond the scope of this work. We note
that the LSTs in the CTA array are designed to handle a stereo trigger rate 
as high as 10 kHz with four times more channels\cite{lst_pmts}. Selecting channels
in the RoI after digitization would also alleviate the dead time.

In any case the increase of discriminator threshold, 
together with the mirror reduction with respect to that of MAGIC
(17 m) will translate to an increase in the energy threshold of the
system. Once the system is triggered, the PMT information needs to  be
recorded. The readout used to digitize and record the information of
the photodetectors is the same as that which is in use in the current MAGIC
system, the Domino Ring Sampler version 4 (DRS4). 

We have run the standard MAGIC analysis program over the simulated
sample  to obtain the sensitivity,
angular and spectral resolutions of the instrument. The analysis
combines the information of the two telescopes at the level of image
parameters (i.e. the so-called Hillas parameters) to obtain a
stereoscopic reconstruction of the shower axis - direction, impact
point and height of the shower maximum ($H_{max}$). Gamma-hadron
separation makes use of the multi-variate event classification method
known as {\it Random Forest}, trained on an statistically
independent sample of MC gamma rays and protons, and fed both with
image parameters and stereoscopic parameters. Energy reconstruction
relies on Look-up tables which correlate the Cherenkov light yield
with the primary gamma-ray energy as a function of various quantities,
mainly the impact parameter, the zenith distance and $H_{max}$. For
more details on the analysis see e.g. \cite{magic2}. 

The gamma/hadron separation cuts, as well as the angular cut around
the nominal gamma-ray direction, have been optimized in order to
maximize the MACHETE sensitivity to point-like sources in each of the
energy bins.

The MC proton numbers have been converted into rates assuming the BESS
measured spectrum\cite{bess}, then scaled up by 20\% to account for
the non-simulated heavier nuclei. For the all-electron spectrum we have
used a parametrization of the {\it Fermi}-LAT measurements between 70 GeV
and 1 TeV\cite{FermiElectrons}.

\subsection{Results of the Montecarlo simulation}

After analysis cuts the energy distribution of the simulated gamma rays 
peaks around 150 GeV for a simulated power-law with photon index $\Gamma$=2.6.

\begin{figure}
\centering
\includegraphics[width=0.8\textwidth]{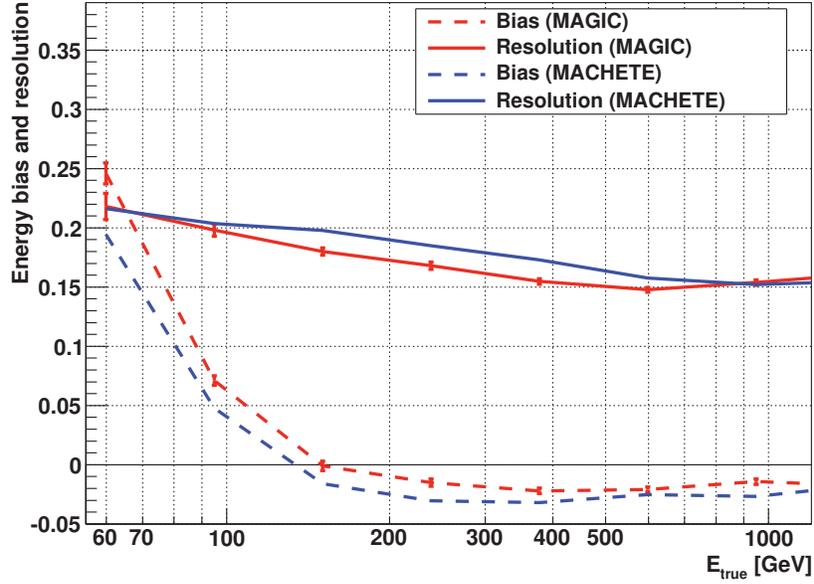}
\caption{Energy resolution and bias in the energy determination of MAGIC and MACHETE 
as a function of $\gamma$-ray true energy.}
\label{fig:energy_resolution}
\end{figure}
We define the energy bias of the system, at a given true energy, as
$E_{bias} = \langle (E_{est}-E_{true})/E_{true} \rangle$ while the
energy resolution is defined as the $\sigma$ of the gaussian fitted
around the peak of the $(E_{est}-E_{true})/E_{true}$ distribution. We
evaluate both parameters using the test sample of MC gamma rays and
compare it to that currently achieved by the MAGIC telescope in figure
\ref{fig:energy_resolution}. The energy resolution is similar for
MAGIC and MACHETE, generally better than 20\%, and reaching 15\% above
1 TeV.

\begin{figure}
\centering
\includegraphics[width=0.8\textwidth]{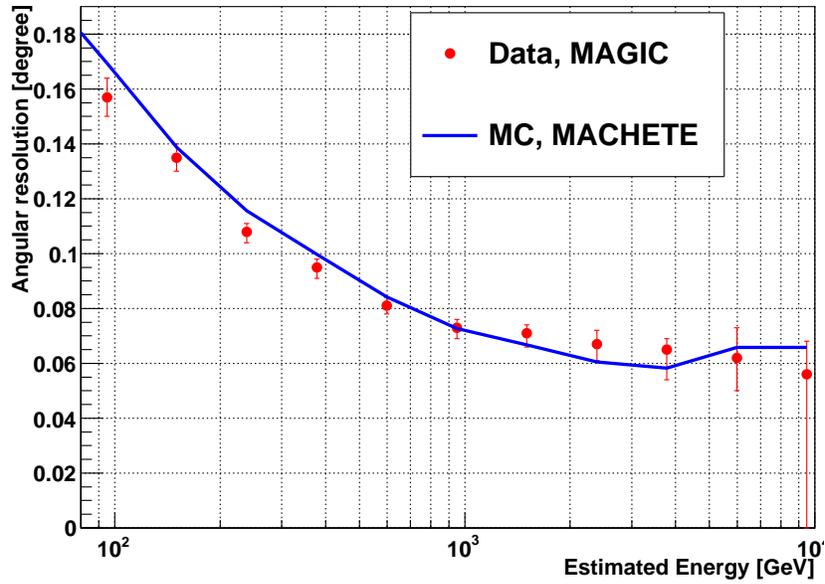}
\caption{Angular resolution of MAGIC and MACHETE, defined as the angle that 
encloses 68\% of the events, as a function of $\gamma$-ray estimated energy.
The source is located 1.6$^\circ$ from the edge of the camera.}
\label{fig:angular_resolution}
\end{figure}
Let us consider the 2-dimensional distribution of reconstructed arrival
directions. We define the angular resolution as the angle that encloses 
68\% of the events. Figure \ref{fig:angular_resolution} shows how it depends
on energy for MC gamma rays and for MAGIC real data.

The fact that the telescopes are fixed to the ground may result in a reduction
of the source position accuracy, which is typically limited by the accuracy and
time stability of the steering and pointing correction systems, but it is
difficult to evaluate the pointing systematics of MACHETE without a more
detailed technical design.

The sensitivity has been calculated using formula 17 of Li and Ma\cite{lima}, 
which is the standard method in
VHE $\gamma$-ray astronomy for the calculation of the significance,
for 5$\sigma$ in 50 h of effective observation time.
The additional conditions number of excess events Nexcess $\geq$ 10, and
Nexcess $\geq 0.05 \times$ the average number of background events are required, to ensure respectively
a minimum gamma-ray statistics and an excess which is safely above the
possible systematic uncertainties in the estimation of the background. 

\begin{figure}
\centering
\includegraphics[width=0.8\textwidth]{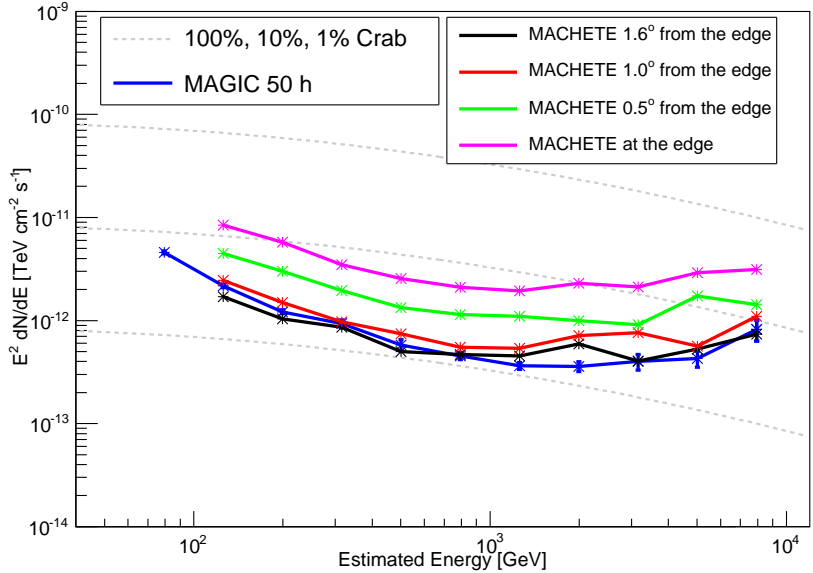}
\caption{Differential sensitivity of MACHETE for a 50 h observation,
  zenith distance $< 30^\circ$, and a source located at different
  angular distances to the edge of the camera. We have used five bins
  per decade of energy. The sensitivity of MAGIC for the same
  observation time and zenith angle range is added for comparison (from
 \cite{magic2}).}
\label{fig:differential_sensitivity_offsets}
\end{figure}

Figure \ref{fig:differential_sensitivity_offsets} shows the sensitivity of MACHETE
averaged on its accessible zenith angle range ($<30^\circ$), a
conventional 50 h observation and a source located at four different
angular distances to the camera edge. If the source is well inside the
camera (i.e. $>$1.0$^\circ$ from the edge) the sensitivity is
within 30\% of that at the center of its FOV.

However, the operation of MACHETE  would be different to that of
existing IACTs. Similar to Milagro or HAWC,  MACHETE is not a pointing
instrument; sources in the accessible range of declinations simply
transit through the MACHETE FOV every day. Since IACT observations are
only possible at night, only the night side of the sky is observable
at a given time of the year. 

If MACHETE were located at the equator it could observe sources in the 
declination range from -30$^\circ$ to +30$^\circ$. This corresponds to
50\% of the sky. We will assume however that MACHETE is located at a 
geographical latitude of 30$^\circ$, similar to the absolute latitude of all the 
existing IACTs. At this latitude the instrument can observe sources 
in the declination range from 0$^\circ$ to +60$^\circ$ (43\% of the sky).

The annual effective observation time of
an IACT system like MAGIC during moonless nights is of around 1000
hours (after technical and bad  weather losses). Using SiPMs may allow
to take 1000 additional hours during Moon time (and in fact MAGIC,
equipped with classical PMT cameras, already records 300 hours of Moon
data every year). However sensitivity and threshold degrade for
Moon observations, hence in what follows we will conservatively stick
to 1000 hours of MACHETE dark observations every year. 
The total transit time of a source is 14 hours at 0$^\circ$
declination, 16 hours at 30$^\circ$ declination and 28 hours at 
60$^\circ$ declination per year.

\begin{figure}
\centering
\includegraphics[width=0.8\textwidth]{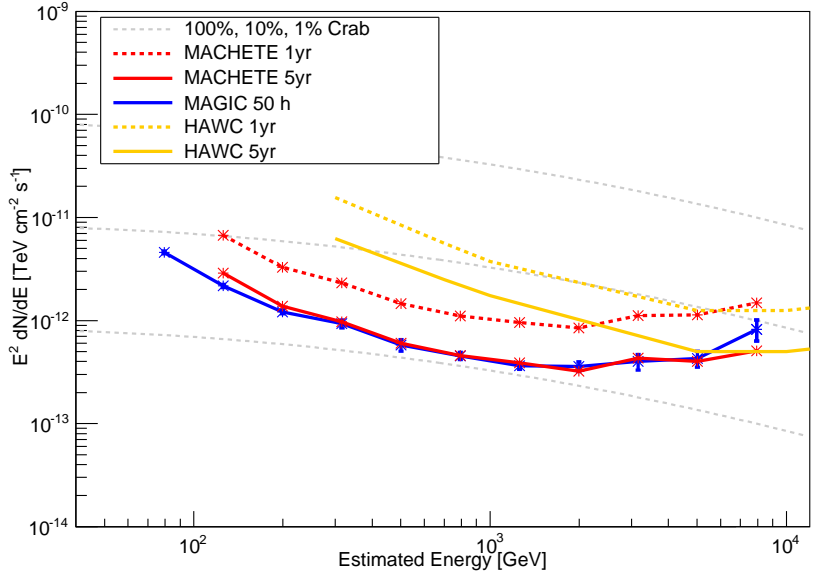}
\caption{Differential sensitivity of MACHETE for a source at 30$^\circ$ declination 
after 1 and 5 years of operation. It is compared to the sensitivity of HAWC for the same
operation time and with the sensitivity of MAGIC for  a dedicated 50-hour observation.}
\label{fig:differential_sensitivity_hawc}
\end{figure}
As the source transits through the FOV of MACHETE, the sensitivity changes. The 
sensitivity for a full transit, or for a whole year of observations, can be obtained 
by integrating the total gamma-ray and background statistics, taking into account the 
change of acceptance as a function of position of the source in the FOV. For this goal 
we have considered only the time a given source actually spends within the {\it optical} 
FOV of MACHETE. Figure
\ref{fig:differential_sensitivity_hawc}  shows the MACHETE sensitivity
calculated in this way for a source at an intermediate declination of
30$^\circ$ after 1 and 5 years of operation. The sensitivity
is compared to that of HAWC\cite{hawc_sensitivity} and that of MAGIC for a dedicated 50-hour
observation. The same definition of sensitivity was used. The best integral sensitivity of MACHETE 
is reached at 500~GeV and 
it is 0.77\% of the Crab flux at 30$^\circ$ declination after 5 years of operation.

The sensitivity of MACHETE is significantly better than the sensitivity of HAWC for the
same observation time and energies below 5~TeV. 
It must be stressed that HAWC has recently come online whilst MACHETE whilst
it may take more than five years to build MACHETE, but even one year of MACHETE 
has better sensitivity than 5~years of HAWC below 2~TeV. In addition, the angular and spectral
resolution would be significantly better than those of HAWC. 

The sensitivity of MACHETE can also be compared to the sensitivity of the planned extragalactic scan
of CTA\cite{CTA_EGAL}. For the full CTA-South array, without divergent pointing, and an observation
of half of the sky for 1000 hours, the expected integral sensitivity of the survey is 
$\sim$0.6\% of the Crab flux above 125~GeV. MACHETE achieves a similar sensitivity in
5 years of operation at a slightly higher energy but for the same fraction of the sky. 

MACHETE would achieve an integral sensitivity of $\sim$12\% of the Crab flux in a single night 
for all sources in the
fraction of the sky that is observable in that specific night. The sensitivity
depends on the declination of the source: it will be about 20\% worse for a source at 
0$^\circ$ declination and 20\% better for a source at 60$^\circ$ declination on account of
the different observation times.
The fraction of the sky covered during one night ranges from around 15\% in Summer to around
20\% in Winter assuming a New Moon night.

In what regards transients MACHETE, HAWC and conventional IACTs have different, 
sometimes complementary, features:
\begin{itemize}
\item In terms of angular resolution MACHETE is similar to conventional IACT and significantly
better than HAWC. This is relevant to localize previously unknown transients.
\item For the same observation time (i.e. in terms of ``instantaneous sensitivity'') 
MACHETE is significantly more sensitive than HAWC and roughly as sensititive as current IACTs. 
It will be however significantly less sensitive than CTA. 
\item The FOV of HAWC is around 5000~sq~deg, i.e. roughly 20 times larger than the FOV of
MACHETE, which is in turn about 10 times larger than the FOV of a conventional IACT. 
\item MACHETE cannot track a source and its daily exposure is restricted to around 20 minutes, while
a source remains inside the FOV of HAWC for several hours and can remain inside the FOV
of a conventional IACT even longer.
\end{itemize}

\section{Physics goals}
\label{Physics}

Surveys covering most of the northern sky have been carried out by the Milagro and Tibet
air shower arrays down to an average sensitivity of 60\% of the Crab flux above 1 TeV. 
HAWC has just started operation with its full array of water tanks and will 
probably reach an integral sensitivity of 100\% of the Crab flux in a day and 
3\% of the Crab flux in five 
years above 1 TeV for much of the northern sky.

Limited surveys of the galactic plane have been performed by IACTs like 
HEGRA, H.E.S.S. and VERITAS. The survey of the inner galaxy carried out by 
H.E.S.S.\cite{hess_survey} 
led to the detection of tens of sources, many of which were unexpected.
Some of them have no obvious counterparts at other wavelengths.
A galactic and an extragalactic survey will probably be performed during the
first years of operation of CTA\cite{cta_surveys}.

At MeV-GeV energies the {\it Fermi}-LAT catalogs have uncovered several classes of sources,
some of which were unexpected before. About one third of the sources remain unidentified.

All in all however most of the extragalactic sky remains unexplored at energies above 100 GeV. 
MACHETE would allow to survey almost half of the sky with a sensitivity of $\sim$0.77\% 
of the Crab flux in 
five years of continuous operation. 
New Active Galactic Nuclei (AGN) will foreseably be discovered. Estimates for 
the number of actual objects that MACHETE could discover in five years differ 
from a handful (based on \cite{cta_surveys}) to several tens (based on \cite{giommi}).

Typically AGNs are observed by IACTs only during bright states in optical, X-rays or VHE.
Instead MACHETE would produce light curves for known and unknown AGNs which 
are not biased by the state of the source. This allows to make correlations with 
other wavelengths and searches for periodicity and delays in VHE emission 
depending on energy. {\it Fermi}-LAT has recently found the first hint for a 
regular cyclic behavior in a $\gamma$-ray emitting blazar (with a timescale of about 2 years, 
\cite{pg1553_periodicity} and the first clear $\gamma$-ray measurement of a delay between flares 
from the gravitationally lensed images of a blazar. This was possible because {\it Fermi}-LAT
makes regular scans of the sky. Both objects have been detected at VHE\cite{gravitational_lens_magic}
and they are bright enough to be detected by MACHETE.

Similar unbiased light curves can be produced for galactic sources. Good examples are the 
gamma ray binaries LSI~+61 303 or LS 5039, which exhibit a wealth of periodic and sporadic 
variability. Such light curves could be used to define data sets with potential signal 
in neutrinos or even gravitational wave experiments. 

MACHETE could prove essential to monitor variable VHE sources for flares and provide triggers to 
instruments at VHE and other wavelengths. The recent detection of a flare with a doubling time 
faster than 4.8 min in the radio galaxy IC 310\cite{ic310_magic} illustrates how little we know
about the temporal behaviour of these VHE sources: it was discovered serendipitiously 
and did not repeat in the following nights or months. The results of an online analysis 
of the data of MACHETE can be available in a matter of a few minutes, allowing steerable IACTs 
like those in CTA to repoint and observe the source. Current IACTs strongly benefit 
from alerts released 
by {\it Fermi}-LAT, but this detector will probably not be available by the time CTA starts 
to take data. 

It is worth to mention again that a camera equipped with SiPMs can be used during moonlight. 
That would allow to double the total observation time in a year and, most importantly, to 
double the duty cycle of the instrument. Moon data, even with degraded performance, would
be highly valuable for monitoring purposes.

As mentioned above other surveys have detected a significant amount of sources that do not
have a clear counterpart at other wavelengths. The survey may identify VHE sources other than 
the already known classes of VHE-emitting AGNs. This may include other active galaxies and other 
extragalactic or galactic objects in general. Some of them may not emit or emit only weakly at 
longer wavelengths (``dark sources'', see \cite{cta_surveys}).

Among other dark sources MACHETE may be able to spot $\gamma$-rays produced by annihilation of
dark matter in clumps or sub-haloes\cite{dm_diemand,dm_brun}. In general the data collected in 
the survey may allow to stack the signal produced in a variety of dark matter $\gamma$-ray-emitting
candidates such as dwarf spheroidals or clusters of galaxies.

Characterization of very extended regions of VHE emission ($>$1~sq~deg) such as the spiral 
arms of our galaxy or the Fermi bubbles\cite{fermi_bubbles} in the sourthern sky, or searches for anisotropy 
of cosmic rays, are probably at the limit of the sensitivity of a five year survey with MACHETE. 

Chances to observe a GRB during its prompt emission (and for minutes before and after) are low, 
although not negligible. The FOV of MACHETE represents 0.7\% of the sky and
the duty cycle of the instrument is 10\%. Assuming a rate of 1 GRB/day, a GRB would fall in the
FOV of MACHETE in average once every four years. However only a small fraction of the GRBs 
would have emission above the energy threshold.

Some of these physics goals are tied to the location of the instrument. Galactic sources are
more numerous in the sourthern sky, so a survey would detect more sources in the south and light
sources of more sources could be produced. Yet, MACHETE may add little to the survey that CTA
will most probably complete in its years of operation. Triggers from extragalactic transients
will be as frequent in both hemispheres. In general it is probably a good idea to locate MACHETE 
at the same location of any of the CTA arrays, so that weather conditions allow to follow 
as many MACHETE triggers as possible with CTA.

For the same cost one may consider to reduce the size of the telescopes and build three or four 
of them. This would improve the sensitivity of MACHETE at the cost of a higher threshold energy.
However many of the goals of this instrument lie in the field of extragalactic physics
where energy threshold is essential. A more detailed technical design and a careful 
cost/benefit analysis would be necessary before settling this question because it is not 
obvious how the performance scales with the cost of the instrument.

\section{Conclusions}

We have described an array of two transit IACTs with a very large FOV of 5$\times$60 sq~deg 
aligned with the meridian. If located at a geographical latitude of 30$^\circ$
it would survey 43\% of the sky (18000 sq.deg) along the year.

Each telescope has a mirror with a spherical shape of 34~m radius of curvature and
a concentric spherical camera. Light concentrators are used to limit the FOV 
of each pixel, so that the effective mirror diameter is 12~m and the corresponding f/D is 1.42. 
This optical setup would result in an optical PSF of 0.06$^\circ$ for an ideal mirror.

The optical and mechanical design of the IACTs is very simple, hence
enabling to reduce the cost significantly compared to conventional steerable telescopes. Its cost is
in fact limited by the cost of the cameras, each of which should be equipped with about 15000 photodetectors.

We have made a full Montecarlo simulation of the instrument using a realistic 
performance for mirror, photodetectors and readout (based on the real performance of these elements
in MAGIC). The instrument has an analysis energy threshold of 150 GeV. 
Even with a smaller mirror the sensitivity of MACHETE for a 50 hour observation is similar to 
that of MAGIC, although it is worth to mention that the QE of the photosensors is significantly higher.

MACHETE however can only observe a source for about 15 hours in a year
although for all sources in 43\% of the sky. For that fraction of the sky MACHETE reaches an 
integral sensitivity of $\sim$0.77\% of the Crab flux after five years of continuous operation. 
For sources observable in a single
night it reaches a sensitivity of $\sim$12\% of the Crab flux. The sensitivity is somewhat 
dependent on the declination due to the different transit speeds.
Angular and spectral resolution are comparable to those of current IACT arrays.

This combination of wide FOV and sensitivity is unprecedented at energies of a few hundreds of GeV. 
It may be essential to alert pointing instruments about sources flaring at VHE in
a good fraction of the observable sky, make surveys of the VHE extragalactic sky and gather unbiased 
light curves of variable VHE sources.

\section*{Acknowledgments}

This paper would have been impossible without the support of the
MAGIC collaboration. 
We gratefully acknowledge the MAGIC collaboration for allowing us to use their Montecarlo 
and data analysis software. We thank J. L. Rasilla (IAC Tenerife), our colleagues at IFAE, 
D. Mazin (ICRR Tokyo), J. Sitarek (U. Lodz), A. Biland (ETH Zurich), R. Paoletti (INFN Pisa) 
and E. Lorenz (MPI Munich) for advice and useful discussions.
Funding for this work was partially provided by the Spanish MINECO under projects
FPA2012-39502, CPAN CSD2007-00042, Multidark CSD2009-00064 and SEV-2012-0234.




\begin{thebibliography}{00}


\bibitem{fermi}
W. B. Atwood et al., The Large Area Telescope on the Fermi Gamma-Ray Space Telescope Mission,
ApJ 697 (2009) 697, 1071.

\bibitem{magic1}
J. Aleksi\'c et al., 
The major upgrade of the MAGIC telescopes, Part I: The hardware improvements and the 
commissioning of the system, 
submitted to Astroparticle Physics and arXiv:1409.6073.

\bibitem{magic2}
J. Aleksi\'c et al., 
The major upgrade of the MAGIC telescopes, Part II: The achieved physics performance using the 
Crab Nebula observations, 
submitted to Astroparticle Physics and arXiv:1409.5594.

\bibitem{hess}
F. A. Aharonian et al., 
Observations of the Crab nebula with HESS. 
Astronomy \& Astrophysics 457 (2006) 899-915.

\bibitem{veritas}
J. Holder et al., Dec. 2008. 
Status of the VERITAS Observatory. 
In: F. A. Aharonian, W. Hofmann, F. Rieger (Eds.), 
American Institute of Physics Conference Series. 
Vol. 1085 of American Institute of Physics Conference Series. pp. 657-660.

\bibitem{milagro}
R. Atkins et al., 
TeV gamma-ray survey of the northern hemisphere sky using the
Milagro observatory, 
ApJ 608 (2004) 680–685

\bibitem{tibet}
M. Amenomori et al., 
A northern sky survey for steady tera-electron volt
gamma-ray point sources using the Tibet Air Shower Array, 
ApJ 633 (2005) 1005–1012.

\bibitem{hawc}
A. Sandoval for the HAWC collaboration,
High Energy Astrophysics with the HAWC Gamma Ray Observatory,
proceedings of RICAP-14, Noto (Sicily, Italy), September 30th- October 3rd, 2014,
available at https://agenda.infn.it/conferenceDisplay.py?confId=7620.

\bibitem{auger}
H. O. Klages et al., 
Optical components for the fluorescence detectors of the Pierre Auger experiment, 
Proc. 27th International Cosmic Ray Conference, Hamburg, id. 764 (2001).

\bibitem{euso}
A. Petrolini, Observation from space of ultra high energy cosmic rays with the
EUSO experiment, Nucl. Phys. B – Proc. Supp. 125 (2003) 212–216.

\bibitem{schliesser}
A. Schliesser, R. Mirzoyan, Wide-field prime-focus imaging atmospheric
Cherenkov telescopes: a systematic study, 
Astrop. Phys. 24 (2005) 382–390.

\bibitem{vassiliev}
V. Vassiliev, S. Fegan, P. Brousseau, Wide-field aplanatic two-mirror telescopes
for ground-based gamma-ray astronomy, 
Astrop. Phys. 28 (2007) 1027.

\bibitem{maccarone}
M.C. Maccarone, P. Assis, O. Catalano et al., Expected performance of the GAW
Cherenkov telescopes array, simulation and analysis. 
Available from arXiv:0707.4352.

\bibitem{mirzoyan}
R. Mirzoyan, M.I. Andersen, 
A 15 deg wide field of view imaging air Cherenkov telescope, 
Astrop. Phys. 31 (2009) 1–5.

\bibitem{lst_pmts}
G. Ambrosi et al., 
LST Technical Design Report, 
CTA internal report MAN-PO/140408, January 2015.

\bibitem{cta_pmts}
T. Toyama et al., 
Evaluation of the basic propertied of the novel 1.5 in. size PMTs from Hamamatsu Photonics and Electron 
Tubes Enterprises, accepted for publications in NIM A.

\bibitem{icrc_sipms}
D. Mazin et al.,
Towards a SiPM camera for current and future generations of Cherenkov telescopes,
Proc. 33rd International Cosmic Ray Conference, Rio de Janeiro, id. 1078 (2013).

\bibitem{Doro}
M. Doro, priv. communication.

\bibitem{mux}
H. Bartko et al., 
Tests of a prototype multiplexed fiber-optic ultra-fast FADC
data acquisition system for the MAGIC telescope,
Nuclear Instruments and Methods in Physics Research A
548 (2005) 464-486.

\bibitem{target}
K. Bechtol et al.,
TARGET: A multi-channel digitizer chip for very-high-energy gamma-ray telescopes
Astroparticle Physics 36 (2012) 156-165.

\bibitem{Paoletti}
R. Paoletti, priv. communication.

\bibitem{Fact}
A. Biland et al., 
Calibration and performance of the photon sensor response of FACT -- The First G-APD 
Cherenkov telescope,
JINST 9	(2014) P10012.

\bibitem{CORSIKA}
D. Heck et al., 
CORSIKA: A Monte Carlo Code to Simulate Extensive Air Showers,
Forschungszentrum Karlsruhe Report FZKA 6019 (1998).

\bibitem{magic_sw0}
T. Bretz, R. Wagner et al.,
The MAGIC Analysis and Reconstruction Software
Proc. 28th International Cosmic Ray Conference, Tsukuba, (2003) p.2947. 

\bibitem{magic_sw1}
S. Lombardi et al., 
Advanced stereoscopic gamma-ray shower analysis with the MAGIC telescopes
Proc. 32nd International Cosmic Ray Conference, Beijing, 3 (2011) 266.

\bibitem{magic_sw2}
R. Zanin et al.,
MARS, the MAGIC analysis and reconstruction software
Proc. 33rd International Cosmic Ray Conference, Rio de Janeiro, id. 773 (2013).

\bibitem{bess}
T. Sanuki et al., 
Precise measurement of cosmic-ray proton and helium
spectra with the BESS spectrometer, 
ApJ 545 (2000) 1135 - 1142.

\bibitem{FermiElectrons}
M. Ackermann et al.,
Fermi LAT observations of cosmic-ray electrons from 7 GeV to 1 TeV,
Phys.Rev.D82 (2010) 092004.

\bibitem{lima}
T.-P. Li and Y.-Q. Ma, 
Analysis methods for results in gamma-ray astronomy,
ApJ 272 (1983) 317.

\bibitem{hess_survey}
F. A. Aharonian et al.,
A new population of very high energy gamma-ray sources in the Milky Way, 
Science 307 (2005) 1938–1942.

\bibitem{cta_surveys}
G. Dubus et al., 
Surveys with the Cherenkov Telescope Array
Astrop. Phys. 43 (2013) 317–330.

\bibitem{giommi}
P. Padovani, P. Giommi,
A simplified view of blazars: the very high energy gamma-ray vision.
MNRAS Letters 446-1 (2015) L41-L45.

\bibitem{pg1553_periodicity}
S. Ciprini, S. Cutini, S. Larsson,
Long--Term Variability of Radio, Optical and Gamma-ray Emission of
PG 1553+113 with the Fermii LAT, 
proceedings of Fifth International Fermi Symposium, October 20-24, 2014,
Nagoya, Japan.

\bibitem{gravitational_lens_fermi}
C. C. Cheung et al.,
Fermi Large Area Telescope detection of gravitational lens delayed $\gamma$-ray flares
from blazar B0218+357, 
ApJ 782 (2014) L14.

\bibitem{gravitational_lens_magic}
R. Mirzoyan et al.,
Discovery of Very High Energy Gamma-Ray Emission From Gravitationally Lensed Blazar 
S3 0218+357 With the MAGIC Telescopes,
ATel 6349, 28 Jul 2014. 

\bibitem{ic310_magic}
J. Aleksi\'c et al., 
Black hole lightning due to particle acceleration at subhorizon scales,
Science, 346/6213 (2014) 1080-1084. 

\bibitem{dm_diemand}
J. Diemand et al.,
Clumps and streams in the local dark matter distribution, 
Nature 454 (2008) 735–738.

\bibitem{dm_brun}
P. Brun et al., 
Searches for dark matter subhaloes with wide-field Cherenkov telescope surveys, 
Phys. Rev. D 83 (1) (2011) 015003.

\bibitem{fermi_bubbles}
M. Su, T. R. Slatyer, D. P. Finkbeiner, 
Giant Gamma-ray Bubbles from Fermi-LAT: Active Galactic Nucleus Activity or Bipolar Galactic Wind?,
ApJ 724 (2010) 1044.

\bibitem{hawc_sensitivity}
A.U. Abeysekara et al., 
Sensitivity of the high altitude water Cherenkov detector to sources of multi-TeV gamma rays,
Astrop. Phys., 50–52 (2013) 26.

\bibitem{CTA_EGAL}
D. Mazin et al.,
Key Science Project Description: Extragalactic Survey,
CTA internal report, SCI-LINK/020714 (version 4.1, Feb. 2015).


\end{thebibliography}


\section*{References}

\end{document}